\documentclass[final]{svjour2}
\usepackage{graphicx}
\usepackage{rotating}
\usepackage{amssymb}
\usepackage{mathptmx}
\usepackage[numbers]{natbib}
\usepackage{color}
\makeatletter
\journalname{Journal of Low Temperature Physics}

\bibpunct{}{}{,}{s}{}{,}

\begin{document}

\newcommand{\hdblarrow}{H\makebox[0.9ex][l]{$\downdownarrows$}-}
\title{Preliminary Heat Capacity and Vapor Pressure Measurements of 2D $^{\mathbf{4}}$He on ZYX Graphite}

\author{S. Nakamura \and K. Matsui \and T. Matsui \and Hiroshi Fukuyama}

\institute{Department of Physics, The University of Tokyo\\ Tokyo, 113-0033, Japan\\
Tel.: +81-3-5841-4193\\Fax: +81-3-5841-4528\\
\email{nakamura@kelvin.phys.s.u-tokyo.ac.jp, hiroshi@phys.s.u-tokyo.ac.jp}}
\date{07.07.2012}

\maketitle


\begin{abstract}

We report preliminary heat capacity and vapor pressure measurements of the first and second layers of $^4$He adsorbed on ZYX graphite.  
ZYX is known to have much better crystallinity than Grafoil, the most commonly-used exfoliated graphite substrate, such as  a ten-times larger platelet size.
This allows us to distinguish different phases in 2D $^4$He much more clearly and may provide qualitatively different insights into this system.
We found a significantly asymmetric density-dependence of the heat-capacity peak associated with the $\sqrt{3}\times\sqrt{3}$ phase formation comparing with that obtained with Grafoil. 
The 2nd-layer promotion density is determined as $11.8 \pm 0.3$~nm$^{-2}$ from the heat-capacity measurement of low density samples in the 2nd layer and vapor pressure measurement.  

\end{abstract}

\section{Introduction}
Helium adsorbed on basal plane graphite has been a fruitful research field for physics of two-dimensional (2D) quantum system. 
Grafoil is the most commonly-used exfoliated graphite in this type of experiments because of its large specific surface area as much as $20$~m$^{2}/$g. 
However, the characteristic length of step-free atomically flat surfaces (platelets) is only about $10-20$~nm. 
This causes the size effects in phase transitions of 2D He system and large amounts of heterogeneous adsorption sites. 
ZYX is an exfoliated graphite with much larger platelet size of $100-200$~nm than Grafoil~\cite{niimi}. 
Previous experiments with the ZYX substrate show much sharper anomalies in heat-capacity and sub-steps in vapor-pressure isotherms at the $\sqrt{3}\times\sqrt{3}$ phase formation in the 1st layer of $^{4}$He~\cite{bretz} and N$_{2}$~\cite{niimi,lt26} than those with Grafoil, respectively. 
In this phase, adsorbed atoms occupy the center of 
every 
one of three adjacent hexagons of carbon honeycomb lattice.
However, there have been few experiments so far with ZYX because of its much smaller surface area ($2$~m$^2/$g) by a factor of ten than Grafoil.

In this report, we show preliminary results of new heat-capacity and vapor-pressure measurements of $^4$He with ZYX.
We measured heat capacities of the 1st layer near the density of the $\sqrt{3}\times \sqrt{3}$ phase ($\rho_{\mathrm{c1}}=6.37$~nm$^{-2}$) and at densities around the 2nd layer promotion where vapor pressure measurements were also carried out.

\section{Experimental}
Details of the experimental setup including the preparation of the ZYX substrate are described in our previous papers~\cite{niimi,lt26}. 
Briefly, the sample cell is made of nylon and contains $19.0$~g of ZYX with a surface area of about 30~m$^2$.
A more detailed description of the determination of surface area is given in the next section. 
Heat capacity is measured using the quasi-adiabatic heat-pulse method with a Cernox CX-1050 thermometer. 
Since the calorimeter is originally designed for measurements at temperatures below $2$~K, it is not thermally isolated sufficiently beyond that temperature. 
Therefore, in this work, we applied constant heat currents to the ZYX substrate in order to avoid too rapid cooling of the temperature of the calorimeter which is sometimes much higher than that of the mixing chamber of the dilution refrigerator. 
The poor thermal isolation limits the maximum waiting time after a heat pulse to a few hundred seconds long which can be shorter than the thermal relaxation time of the system. 
We found that, by multiplying our data at $\rho=\rho_{\mathrm{c1}}$ by a $T$-linear correction factor $\eta$ from 1.18 to 1.47 in the temperature range of 2.6 $\leq T \leq$ 3.2 K, we obtain a very good agreement with those by Bretz~\cite{bretz} within scatterings (5\%) of the two data sets. 
In principle, $\eta$ should depend only on temperature and such an assumed $T$-linear dependence of $\eta$ gives $\eta=1$ at $T\approx 2.2$~K where we know the correction is unnecessary because of much better thermal isolation. 
Therefore we applied this correction to our all $C$ data at the temperatures. 
The addendum heat capacity in this temperature range is rather large being about 40\% of $C_{\mathrm{peak}}$. 
Vapor pressure was measured by a capacitive-pressure-gauge with a 0.2-mm-thick hard-silver diaphragm located on the mixing chamber plate.
The sensitivity is $1\times10^{-3}$~mbar. 

\section{Results and Discussion}
\subsection{The 1st layer $\sqrt{3}\times \sqrt{3}$ phase}
\begin{figure}[htbp]
\begin{minipage}{0.49\linewidth}
\begin{center}
\includegraphics[%
  width=0.98\linewidth,
  keepaspectratio]{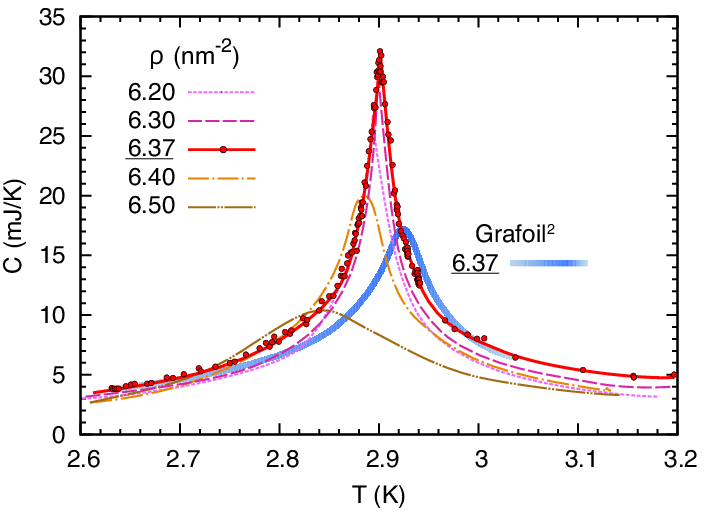}
\caption{(Color online) Heat capacity of the 1st layer of $^4$He on ZYX at densities in the vicinity of the $\sqrt{3}\times\sqrt{3}$ phase at $\rho = 6.37$~nm$^{-2}$. Thick line represents the data with Grafoil~\cite{bretz}.}
\label{root3qfs}
\end{center}
 \hfill
\end{minipage}
\hspace{0.02\linewidth}
\begin{minipage}{0.49\linewidth}
\begin{center}
\includegraphics[%
  width=0.98\linewidth,
  keepaspectratio]{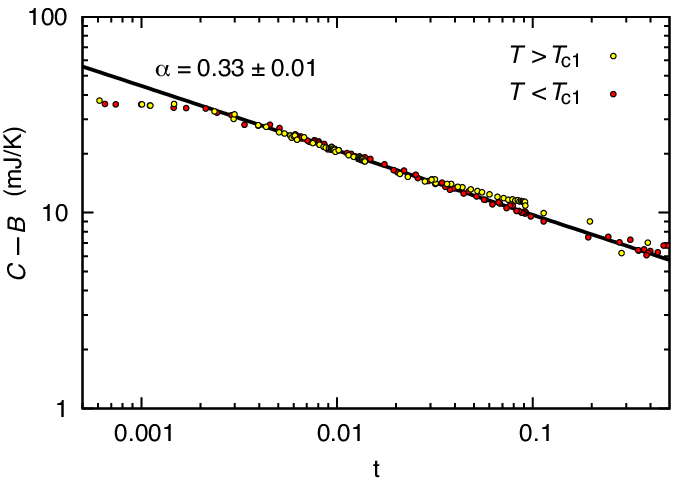}
\caption{(Color online) Measured heat capacities plotted as a function of reduced temperature $t$ for the 1st-layer $\sqrt{3}\times\sqrt{3}$ phase ($\rho = 6.37$~nm$^{-2}$) of $^4$He on ZYX. A constant background $B=-6.1\pm0.5$ has been subtracted from the raw data. The critical exponent $\alpha$ is deduced as $0.33\pm 0.01$.}
\label{1by3loglog}
\end{center}
\end{minipage}
\end{figure}

Figure~\ref{root3qfs} shows measured heat capacities of the 1st layer at densities in the vicinity of $\rho_{c1}$.
The anomaly is much sharper, with a two-times larger peak-height, than that obtained in the previous experiment \cite{greywall} with Grafoil shown as a thick line in the figure. 
The melting transition of the $\sqrt{3}\times \sqrt{3}$ phase is explained by the three-state Potts model~\cite{potts} which gives a symmetrical specific heat about $T = T_{c1}$ with a critical exponent  $\alpha$ of $\frac{1}{3}$. 
These behaviors are clearly seen in Fig.~\ref{1by3loglog} where our data give $\alpha=0.33\pm 0.01$. 
Here, a constant background heat capacity $B=-6.1$~mJ/K is subtracted. 
This $\alpha$ value is consistent with the theoretical expectation and agrees with $\alpha=0.35-0.38$ obtained in the previous experiment with ZYX~\cite{bretz}.
The data deviate systematically from the fitting at $(T-T_{\mathrm{c1}})/T_{\mathrm{c1}}\geq 0.03$, which is not seen in the previous measurement where a temperature-dependent $B$ is subtracted~\cite{bretz}. 
The model also requires an entropy change of $k_{\mathrm{B}}\ln 3$ per particle associated with the phase transition. 
Our corrected data are consistent with this expectation. 

\begin{figure}[htbp]
\begin{center}
\includegraphics[%
  width=0.83\linewidth,
  keepaspectratio]{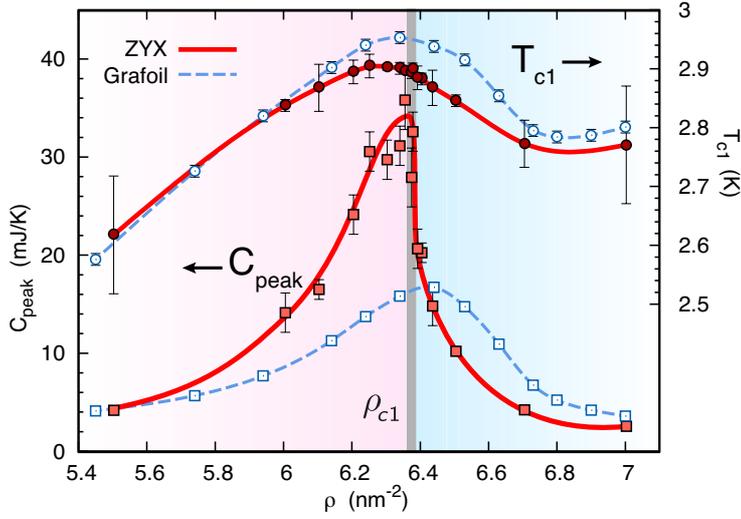}
\caption{(Color online) Heat-capacity peak height $C_{\mathrm{peak}}$ on the left axis and peak temperature $T_{\mathrm{c1}}$ on the right axis of  $^4$He on ZYX (solid symbols; this work) and Grafoil~\cite{greywall} (open symbols) at densities around $\rho_{c1}$. The peak heights of the latter data are normalized to be consistent with the surface area of this work.}
\label{C1by3}
\end{center}
\end{figure}

The maximum peak height is expected to appear at $\rho=\rho_{c1}$, which allows us to determine the surface area of ZYX to be $30.5 \pm 0.2$~m$^{2}$ assuming that the perfect $\sqrt{3}\times\sqrt{3}$ structure is formed there.
Using this value and the $^4$He gas amount introduced into the calorimeter, we can calculate the areal density $\rho$. 
The surface area determined in this way is in good agreement with the value ($= 30.7$~m$^{2}$) obtained in the adsorption isotherm measurement~\cite{lt26}. 
The significantly asymmetric density dependence of $C_{\mathrm{peak}}$ is shown in Fig.~\ref{C1by3} comparing with  the result of Grafoil~\cite{greywall}. 
$C_{\mathrm{peak}}$ with ZYX decreases precipitously by half within $0.05$~nm$^{-2}$ above $\rho_{c1}$ though it decreases only $5\%$ over $0.12$~nm$^{-2}$ below $\rho_{c1}$. 
It means that the excess atoms easily break the commensurate symmetry unlike holes (particle-hole asymmetry). 
The data with Grafoil show a much broader and more symmetric maximum around $\rho_{c1}$ over $\pm 0.2$~nm$^{-2}$, which is presumably due to a much larger amount of surface heterogeneities in this substrate. 
The usage of the higher quality substrate in this experiment will provide further information on the so-called {\it commensurate-incommensurate} or {\it domain-wall} density region ($\rho>\rho_{c1}$) in the near future. 

\subsection{The 2nd-layer promotion}
The experimental determination of the 2nd-layer promotion density $\rho_{1\rightarrow2}$ is important because it is a calibration point in determining density scale for experimentalists who use exfoliated graphite substrates which necessarily contain finite amounts of surface irregularities and in testing accuracies of calculation for theorists who use different computational methods or assumptions. 

\begin{figure}[htbp]
\begin{center}
\includegraphics[%
  width=0.53\linewidth,
  keepaspectratio]{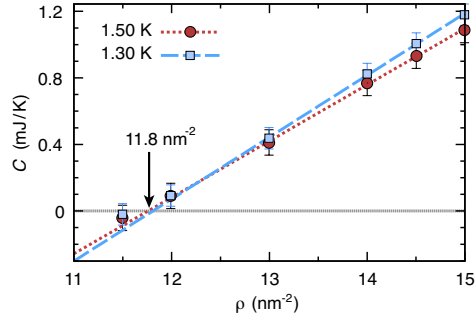}
\caption{(Color online) Heat capacity isotherms of the 2nd layer of $^4$He on ZYX graphite. The lines are linear fittings of the data of the four highest densities.}
\label{iso}
\end{center}
\end{figure}

We made heat capacity measurements of $^4$He on ZYX at densities of $11.5<\rho<15$~nm$^{-2}$ in the temperature range of $1.3<T<1.7$~K where the total heat-capacity is expectedly dominated by a contribution from the Dulong-Petit type $T$-independent and density-linear heat capacity of the 2nd-layer gas. 
We estimated here a heat capacity contribution from the 1st-layer solid as a Debye heat-capacity with 
$\Theta_{D}=56-58$~K, which is derived from the known $\rho$-dependency of the first layer density~\cite{rogerjltp1998} and the Gr\"{u}neisen parameter for $\Theta_{D}$~\cite{ecke},
  and subtracted it from the total heat capacity. 
As expected, the measured heat-capacity isotherms at five different temperatures ($T=1.3,$ $1.4,$ $1.5,$ $1.6,$ $1.7$~K) follow straight lines very well, which gives $\rho_{1\rightarrow2}=11.8\pm 0.3$~nm$^{-2}$ as shown in Fig.~\ref{iso} (only two of them are plotted here). 
A similar $\rho_{1\rightarrow2}$ value ($=12.0$~nm$^{-2}$) is reported  previously~\cite{greywall} based on the same analysis, where the density grid is not fine enough and Grafoil substrate is used. 
Our value is slightly higher than $11.2$~nm$^{-2}$ determined from a neutron scattering experiment~\cite{rogerjltp1998}. 

\begin{figure}[htbp]
\begin{center}
\includegraphics[%
  width=0.62\linewidth,
  keepaspectratio]{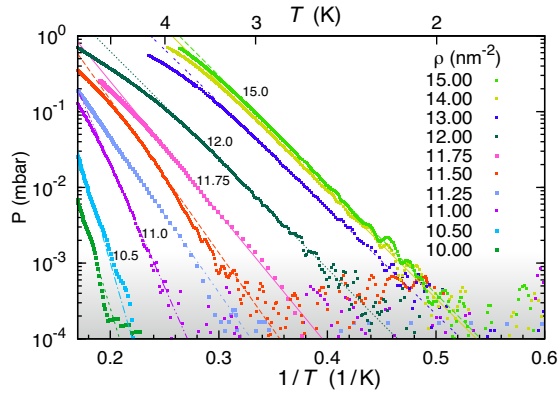}
\end{center}
\caption{(Color online) Temperature dependence of vapor pressure of $^4$He on ZYX graphite at densities below and above the 2nd layer promotion ($\approx 12$~nm$^{-2}$). The numbers denote total $^4$He coverages in nm$^{-2}$. }
\label{demo2}
\end{figure}

Besides heat capacity measurements, we also measured vapor pressure $P$ of samples at selected densities around $\rho_{1\rightarrow2}$ as a function of temperature as shown in Fig.~\ref{demo2}. 
From these data, we can deduce isosteric heat $q_{st}$ for each density using the following equation: 
\begin{equation}
q_{\mathrm{st}}=k_{\mathrm{B}}T^2\left(\frac{\partial \ln P}{\partial T}\right)_{\rho}.
\end{equation}

\begin{figure}[htbp]
\begin{minipage}{0.49\linewidth}
\begin{center}
\includegraphics[%
  width=0.98\linewidth,
  keepaspectratio]{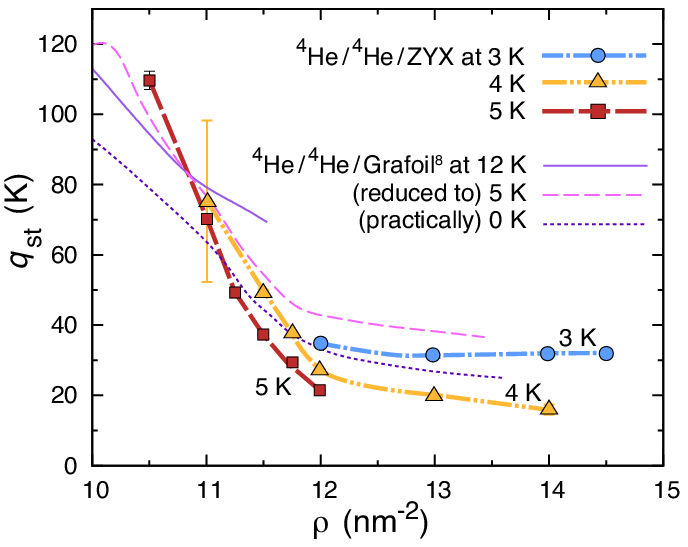}
\caption{(Color online) Isosteric heat of $^4$He deduced from the measured vapor pressure at densities near the 2nd layer promotion. The data points represent our data with ZYX graphite and the thick lines are guides to the eye. The thin lines are data with Grafoil by other workers~\cite{elgin}. }
\label{qst}
\end{center}
\hfil
\end{minipage}
\hspace{0.02\linewidth}
\begin{minipage}{0.49\linewidth}
\begin{center}
\includegraphics[%
  width=0.98\linewidth,
  keepaspectratio]{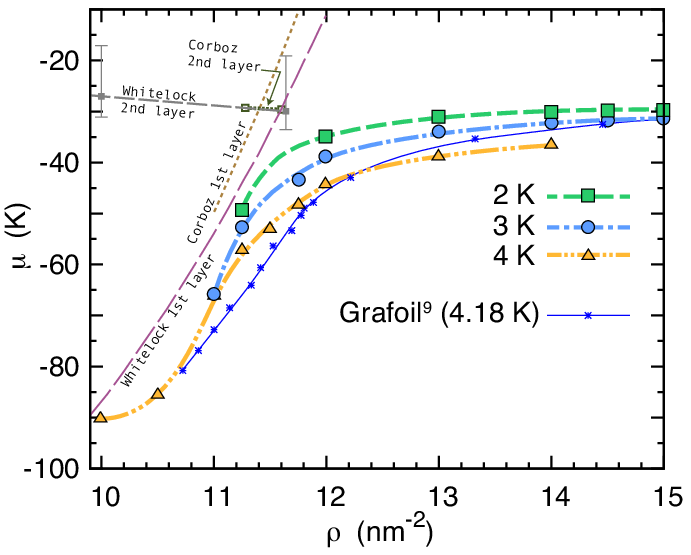}
\caption{(Color online) Chemical potential of $^4$He film on ZYX deduced from the measured vapor pressure. The lines through the data points are guides to the eye. Data with Grafoil by other workers~\cite{polanco} are also shown. The dotted lines represent theoretical calculations of the 1st and 2nd layers of  $^4$He on graphite by Corboz {\it et al.}~\cite{corboz} and dashed lines by Whitlock {\it et al.}~\cite{whitlock}}
\label{mu}
\end{center}
\end{minipage}
\end{figure}

\noindent The result is shown in Fig.~\ref{qst}. 
$q_{st}(\rho)$ seems to have a kink at a density slightly lower than 12~nm$^{-2}$, above which it is almost density-independent at $T =$ 3 K. 
Apparently, the kink corresponds to $\rho_{1\rightarrow2}$. 
By taking more data at densities in between 11 and 12~nm$^{-2}$ using a pressure gauge with a much higher sensitivity than the present one, we should be able to locate $\rho_{1\rightarrow2}$ more precisely from pressure measurement in the near future. 
Figure~\ref{qst} also shows data taken at much higher temperatures with Grafoil by previous workers~\cite{elgin}.
The weak density dependence even for the zero-temperature behavior estimated from their data would indicate the larger amount of surface heterogeneities in their substrate. 

The vapor pressure measurement also gives a different insight into the adsorbed film system.
Chemical potential $\mu$ of $^4$He gas in the sample cell should be equal to that of the 2D $^4$He film in thermal equilibrium with the gas. $\mu$ can be estimated from $P$ as
\begin{equation}
\mu=k_{\mathrm{B}}T\ln\left[\frac{P}{k_{\mathrm{B}}T}\left(\frac{h^2}{2\pi m k_{\mathrm{B}} T}\right)^{\frac{3}{2}}\right]
\end{equation}
on the assumption of ideal gas. 
The resultant $\mu$ values at several fixed temperatures are plotted in Fig.~\ref{mu}. 
The data show again a kink structure at $\rho_{1\rightarrow2}$ at lower temperatures. 
The overall behavior is similar to that with Grafoil by previous workers~\cite{polanco}. 
The chemical potential $\mu_2$ of the 2nd layer ($\rho>$12~nm$^{-2}$) is nearly density independent at $\approx -30$~K. 
In addition, this value is in good agreement with existing theoretical calculations by Corboz {\it et al.}~\cite{corboz} denoted by a horizontal dotted line and by Whitlock {\it et al.}~\cite{whitlock} denoted by a horizontal dashed line in the figure. 
On the other hand, the chemical potential $\mu_1$  of the 1st layer ($\rho<$12~nm$^{-2}$) seems to be consistently lower than the same theoretical calculations. 
However, this may simply be a finite temperature effect.
It is desirable to extend the present pressure measurements down to much lower temperatures using a higher sensitivity pressure gauge. 

\section{Conclusions}
By using the ZYX substrate, which is an exfoliated graphite with better crystallinity than Grafoil, we obtained preliminary heat-capacity data for the first and second adsorbed layers of $^4$He. 
Because of the poor thermal isolation of our calorimeter from the surroundings in the temperature range of 2.6 $\leq T \leq$ 3.2 K, we had to apply relatively large corrections ($1.18-1.47$) to the measured heat capacities. 
Nevertheless, we could observe two characteristic features which demonstrate the better surface quality of our substrate.
First, the heat-capacity anomaly associated with the $\sqrt{3}\times\sqrt{3}$ phase formation in the 1st layer is much sharper than that with Grafoil. 
Second, we found a significantly asymmetric density-dependence of the heat-capacity peak of this phase, which is the first observation providing a new insight into the commensurate-incommensurate transition. 
The 2nd-layer promotion density is determined as $11.8\pm 0.3$~nm$^{-2}$ by extrapolating the linear density-dependences of the heat capacity isotherms of the 2nd layer.
This value is consistent with our own vapor pressure measurement and with the existing theoretical calculations. 
\begin{acknowledgements}
This work was financially supported by Grant-in-Aid for Scientific Research (A) (No. 2222440042) and on Priority Areas (No. 17071002) from MEXT, Japan. S.N. acknowledges support from the Fuuju-kai Fellowship. S.N. and K.M. also acknowledge support from the Global COE Program ``the Physical Sciences Frontier'', MEXT, Japan.\end{acknowledgements}




\end{document}